\documentclass[twocolumn,amsmath,amssymb,aps,prd,10pt,superscriptaddress,nofootinbib,noeprint,preprintnumbers,floatfix]{revtex4-2}
\pdfoutput=1

\usepackage{graphicx, color}

\usepackage[letterspace=-10]{microtype} 

\usepackage{bm, amsmath, amsfonts}
\usepackage{multirow, tabularx, dcolumn}
\usepackage{mathtools} 
\usepackage{booktabs}
\usepackage{placeins}
\usepackage{soul} 

\usepackage[utf8]{inputenc} 
\usepackage{hyperref}

\usepackage{xcolor}
\definecolor{jlab_red}{RGB}{192,39,45}
\definecolor{jlab_orange}{RGB}{249,102,0}
\definecolor{jlab_blue}{RGB}{47,122,121}
\definecolor{jlab_green}{RGB}{65,125,10}
\definecolor{jlab_grey}{RGB}{125,125,125}

\definecolor{dstarpi_s}{RGB}{191,39,45}
\definecolor{dstarpi_d}{RGB}{65,125,10}
\definecolor{dstarpi_mix}{RGB}{248,102,0}
\definecolor{dpi_d}{RGB}{51,92,129}

\hypersetup{%
pdftitle = {eta and eta′ production in J/psi radiative decays from quantum chromodynamics},
pdfsubject = {QCD},
pdfkeywords = {QCD, Hadron, Charmonium, Physics, Lattice, Meson},
pdfauthor = {Hadron Spectrum Collaboration},
colorlinks = {true},
filecolor = {black},
linkcolor = {jlab_blue},
menucolor = {black},
citecolor = {jlab_green},
urlcolor = {jlab_green},
}{}

\begin{document}

\preprint{JLAB-THY-25-4368}

\title{$\eta$ and $\eta'$ production in $J/\psi$ radiative decays from quantum chromodynamics}

\author{Mischa~Batelaan}
\email{mbatelaan@wm.edu}
\affiliation{Department of Physics, College of William and Mary, Williamsburg, VA 23187, USA}
\author{Jozef~J.~Dudek}
\email{dudek@jlab.org}
\affiliation{Department of Physics, College of William and Mary, Williamsburg, VA 23187, USA}
\affiliation{\lsstyle Thomas Jefferson National Accelerator Facility, 12000 Jefferson Avenue, Newport News, VA 23606, USA}
\author{Robert~G.~Edwards}
\email{edwards@jlab.org}
\affiliation{\lsstyle Thomas Jefferson National Accelerator Facility, 12000 Jefferson Avenue, Newport News, VA 23606, USA}

\collaboration{for the Hadron Spectrum Collaboration}
\noaffiliation

\date{ June 11, 2025 }

\begin{abstract}
\noindent
We present a first principles calculation within quantum chromodynamics (QCD) of the radiative decays of the $J/\psi$ into the light pseudoscalar mesons $\eta$ and $\eta'$. Within a lattice computation we obtain the transition form--factors as a function of photon virtuality from the timelike region, accessible experimentally via the `Dalitz' decay $J/\psi \to e^+ e^-\, \eta^{(\prime)} $, through to the real photon point corresponding to $J/\psi \to \gamma\, \eta^{(\prime)} $.
This is the first calculation in lattice QCD with two (heavier than physical) degenerate flavors of light quark and a heavier strange quark, in which the $\eta'$ appears as the \emph{first--excited} state with pseudoscalar isoscalar quantum numbers. We access it reliably by using variationally \emph{optimized operators}, use of which also improves the purity of the $J/\psi$ and $\eta$ signals, reducing systematic uncertainties. 
High quality results at a large number of kinematic points are obtained in a typically noisy disconnected process by using a novel correlator averaging procedure.
Our results show the expected enhanced production of the $\eta'$ over the $\eta$ in this process, and suggest that the demonstrated lattice technology is suitable for future calculations considering processes in which light meson \emph{resonances} are produced.
\end{abstract}

\maketitle

\noindent \emph{Introduction} --- 
Charmonium radiative decays provide a tool to access a range of light meson systems through a common and, in principle, relatively straightforward production process. 
They are of particular contemporary interest owing to the huge BESIII dataset of $J/\psi$ and $\psi(2S)$ radiative decays, from which high quality signals for light--quark resonances have been extracted~\cite{
BESIII:2015rug, BESIII:2018ubj, Rodas:2021tyb, Sarantsev:2021ein}. This process has long been considered important in searches for \emph{glueballs} by virtue of it being `glue-rich' owing to the fact that the charm--anticharm pair in the $J/\psi$ must annihilate to gluons before producing the light meson system.
A recent examination of BESIII data on the $J/\psi \to \gamma \, \eta \eta'$ process has presented evidence~\cite{BESIII:2022riz} for an exotic $J^{PC}=1^{-+}$ resonance, the $\eta_1(1855)$, which is the first experimental candidate for an isoscalar \emph{hybrid}, a meson in which the gluonic field plays a vital role.

The simplest radiative decay processes are those leading to a single relatively long--lived hadron, with the largest rates being for $J/\psi \to \gamma\,  \eta$ and $J/\psi \to \gamma\,  \eta'$, the branching fractions for which are measured at the per--cent level of precision by BESIII~\cite{BESIII:2023fai}.
There are sufficiently many $J/\psi$ events in the BESIII dataset that the further suppressed processes $J/\psi \to e^+ e^-\, \eta^{(\prime)}$, where a timelike virtual photon produces the fermion pair, have also been measured~\cite{BESIII:2018qzg, BESIII:2018aao}.

\smallskip

The importance of this production process for light meson spectroscopy demands consideration within first--principles QCD, and its non--perturbative nature with the presence of multiple scales suggests that the general--purpose tool of lattice QCD is probably most appropriate. As described below, such calculations will be highly challenging.
Recent first attempts to compute $J/\psi \to \gamma \eta^{(\prime)}$~\cite{Jiang:2022gnd, Shi:2024fyv} opted to work with only one or two flavors of light quark, and as such do not have a true distinction between the $\eta$ and the $\eta'$, something which is only present in a three--flavor version of QCD. In such a theory, with strange quarks that are not degenerate with the light quarks, the $\eta'$ appears as the \emph{first excited} state in the pseudoscalar isoscalar channel, lying above the ground--state $\eta$.

The usual approach in lattice QCD of evaluating correlation functions at large time--separations where ground--states dominate will give access only to the $\eta$ and not the $\eta'$, and as such we require an \emph{optimized operator} that has dominant overlap with the $\eta'$. We are able to construct this by variational analysis in a basis of operators with pseudoscalar isoscalar quantum numbers~\cite{Dudek:2011tt,Dudek:2013yja}. 

The large mass gap between charmonium mesons like the $J/\psi$ (above 3 GeV) and the light mesons (around 1 GeV) means that in order to access the region of photon virtuality, $Q^2$, around zero, corresponding to the radiative decay, we need hadrons to carry large values of the discrete three--momentum allowed in the finite spatial volume defined by the lattice.

The $c\bar{c}$ annihilation in this process causes it to be described by a \emph{disconnected} diagram, and these are generally found to be noisy in lattice QCD calculations, requiring consideration of techniques that can improve the signal.

In this calculation we will resolve all these challenges, using optimized operators for hadrons, independently determined at many momenta.
\emph{Distillation}~\cite{HadronSpectrum:2009krc} will be used, allowing efficient computation of a huge number of correlation functions, many of which are related by exact and approximate symmetries in a way which allows them to be averaged over, improving signal quality.

We obtain the transition form--factors for $\eta$ and $\eta'$ production with good statistical precision at 70 discrete values of $Q^2$, and describe the virtuality dependence using appropriate parameterizations. The real--photon decay rates are determined and found to take values some way below the experimentally measured rates. 
Establishing that signals of sufficient quality can be obtained in this sector, opens up future calculations that will consider processes like $J/\psi \to \gamma\,  (\pi\pi, K\bar{K})$ where interesting light meson resonances appear. The required finite--volume tools needed in this case exist  and have been applied in other, simpler, cases~\cite{Briceno:2021xlc, Briceno:2015dca, Briceno:2016kkp, Radhakrishnan:2022ubg, Ortega-Gama:2024rqx, Alexandrou:2018jbt, Leskovec:2025gsw}.

More details of the work described in this letter can be found in an associated write-up.

\medskip
\noindent\emph{Calculation} ---
The lattice used in this calculation has $2+1$ flavors of dynamical quarks, with anisotropic lattice spacings of $a_s\approx 0.12\, \textrm{fm}$, $a_t^{-1} \approx 5.7\, \mathrm{GeV}$ on a volume of $(L/a_s)^3\times (T/a_t) = 20^3\times 256$. The strange quark mass is tuned to approximately its physical value, while the degenerate light quarks are such that $m_\pi \sim 391\, \mathrm{MeV}$. For the calculation of all correlation functions, on an ensemble of 288 configurations, the distillation approach \cite{HadronSpectrum:2009krc} is used with $N_{\textrm{vec}}=128$. This lattice and related lattices with other volumes have been previously used to determine light meson spectra~\cite{Dudek:2010wm,Dudek:2011tt,Dudek:2013yja, Dudek:2016cru, Briceno:2017qmb, Woss:2019hse}, charmonium spectra~\cite{HadronSpectrum:2012gic, Wilson:2023hzu, Wilson:2023anv}, and charmonium radiative transitions~\cite{Delaney:2023fsc}.

The radiative decays $J/\psi \to \gamma \, \eta^{(\prime)}$ can be described by a matrix element of the electromagnetic current featuring a single Lorentz--invariant form--factor,
\begin{align}
  \label{eq:4}
  \begin{split}
    &\big\langle \eta^{(\prime)}(\mathbf{p}') \big| j_{\mathrm{em}}^{\mu}(0) \big| J/\psi(\mathbf{p}, \lambda) \big\rangle =  \\
    &\qquad \qquad \qquad \epsilon^{\mu\nu\rho\sigma}\, p'_{\nu}\, p_{\rho}\, \epsilon_{\sigma}(\mathbf{p}, \lambda) \, F_{\psi \eta^{(\prime)}}(Q^2) \, ,
  \end{split}
\end{align}
where $\epsilon_{\sigma}(\mathbf{p},\lambda)$ is the polarization vector describing the $J/\psi$ with helicity $\lambda$. The form--factor at zero photon virtuality can be related to the partial decay width
\begin{equation*}
\Gamma \big(J/\psi \to \gamma \, \eta^{(\prime)} \big) = \frac{4}{27} \alpha \, |\mathbf{q}|^3 \, \big|F_{\psi \eta^{(\prime)}}(0)\big|^2 \, ,
\end{equation*}
with $|\mathbf{q}|$ being the photon momentum in the rest--frame of the $J/\psi$.

To interpolate mesons efficiently, and to access the excited--state $\eta'$, we use optimized operators in the form of a linear combination of operators in a basis of fermion bilinears, obtained by solving variationally a matrix of two-point correlation functions~\cite{Michael:1985ne, Luscher:1990ck, Dudek:2007wv, Dudek:2011tt, Dudek:2013yja}. This is done independently at each momentum, in each irreducible representation (\emph{irrep}) of the (boosted) lattice symmetry, and the so--determined energies for the $J/\psi$, $\eta$ and $\eta'$ are all consistent with a continuum--like relativistic dispersion relation with an anisotropy, ${\xi = 3.51^{+0.08}_{-0.06}}$.

To a very good approximation, the radiative decay processes considered in this letter are dominated by the photon coupling only to the charm quark in the $J/\psi$, and as such we will compute only the charm part of the electromagnetic current\footnote{Justification for this can be found in the associated longer write-up.}. The vector current on this anisotropic lattice must be renormalized and to be consistent with the $\mathcal{O}(a)$--improvement of the charm quark action, requires addition of a tree--level improvement term. The multiplicative renormalization is set by using the $\eta_c$ form--factor at $Q^2=0$ determined in a previous work with the same lattice action~\cite{Delaney:2023fsc}.

The matrix--elements we need appear in three--point correlation functions that we will construct using optimized meson operators on fixed timeslices ($0, \Delta t$) with the current inserted at all times between them, $ 0 \le t \le \Delta t$,
\begin{equation*}
  \label{eq:1}
  C(t, \Delta t) =  \big\langle 0 \big|  \Omega_{\eta^{(\prime)}}(\Delta t) \; j(t) \; \Omega^{\dagger}_{\psi}(0) \big| 0 \big\rangle \, .
\end{equation*}
Dividing out the ground--state time dependence from this gives
\begin{equation*}
  \tilde{C}(t, \Delta t) \equiv \tfrac{\sqrt{2 E_{\eta^{(\prime)}} \, 2 E_\psi} }{ e^{-E_{\eta^{(\prime)}} (\Delta t - t)} e^{- E_\psi t} } \,  C(t, \Delta t) = \langle \eta^{(\prime)} | j(0) | \psi \rangle + \ldots \, ,
\end{equation*}
which is the desired matrix--element up to excited--state corrections that will be suppressed owing to the use of optimized operators.

To improve the signal--to--noise we calculate all three--point correlation functions corresponding to each $Q^2$ value, meaning all momentum directions for the mesons and all rows of all irreps, while fixing the direction of the momentum inserted at the current.
For the $J/\psi$ we consider optimized operators with momenta $|\mathbf{p}|^2\le 4 \left(\tfrac{2\pi}{L}\right)^2$, while for the $\eta$ and $\eta'$ we have $|\mathbf{p}'|^2\le 6 \left(\tfrac{2\pi}{L}\right)^2$, and for the current insertion momenta $|\mathbf{q}|^2\le 16 \left(\tfrac{2\pi}{L}\right)^2$.
The resulting combinations lead to 1748 non--zero correlation functions, but suitable averaging of correlators at the same value of $Q^2$ allows us to reduce this down to 70 well--determined values of $F(Q^2)$ for each of $\eta$ and $\eta'$. Our averaging procedure is in two stages: the first stage averages correlators which are related by exact symmetries of the (boosted) lattice, using lattice Clebsch-Gordan coefficients~\cite{Dudek:2012gj} in the Wigner-Eckart theorem to bring them into a common form. These averaged correlators have their time--dependence fitted with a form
\begin{equation*}
  \label{eq:threePtFit}
  \tilde{C}(t, \Delta t) = J + A_{\text{src}}\,  e^{-\delta E_{\text{src}} t}   + A_{\text{snk}} \, e^{-\delta E_{\text{snk}} ( \Delta t - t )} \, ,
\end{equation*}
where $J$ is the desired matrix--element, and $A_{\text{src}},\, A_{\text{snk}}$, $\delta E_{\text{src}},\, \delta E_{\text{snk}}$ are free parameters which capture any remaining excited--state contributions. Such fits are performed over a range of time--windows with the results combined into a conservative average using the Akaike information criterion (AIC) based weighting procedure presented in Ref.~\cite{Jay:2020jkz}.

\begin{figure}
  \centerline{\includegraphics[width=.99\columnwidth]{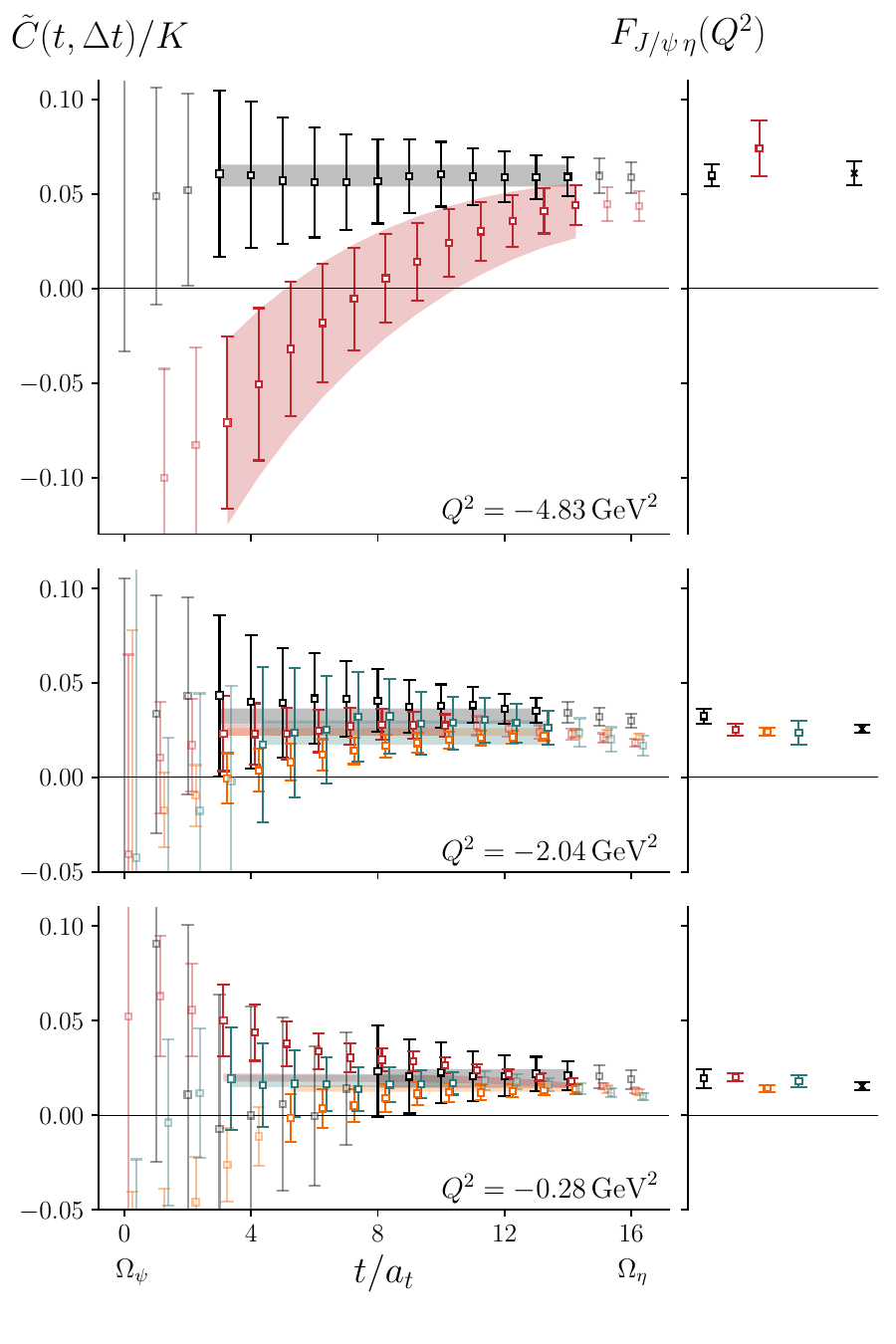}}
  \vspace*{-8mm}
  \caption[]{\label{fig:corrs}
    Examples of timeslice fitting and correlator averaging procedure at three values of $Q^2$ for $J/\psi \to \gamma\,\eta\,$ as described in the text.
  }
\end{figure}

The second stage of averaging relies upon the Lorentz symmetry assumed in Eqn.~\ref{eq:4}. We propose that the discretization on this lattice is such that this symmetry is manifested approximately, and that computations using different irreps are related by  straightforward \emph{subduction}~\cite{Thomas:2011rh}. Constructing an over--constrained linear least--squares fit to matrix--elements with just a single assumed $F(Q^2)$ implements our averaging.

Figure~\ref{fig:corrs} illustrates the time--dependence fitting and the subsequent averaging at three different values of $Q^2$ for $\eta$ production matrix--elements. The $\Delta t/a_t = 16$ correlators in each of the left panels show all possible combinations of irreps at each momentum value, already averaged using the Wigner-Eckart theorem, with the bands showing the single best timeslice fit over many time--windows (as measured by AIC). The right hand panels show the values of the form--factor extracted from each individual correlator, seen to be in close agreement, supporting our hypothesis of an approximate Lorentz symmetry, while the rightmost point gives the result of the over--constrained linear least-squares fit to these, which serves as our final estimate at that $Q^2$.
Results presented in the remainder of this letter follow from using spatially--directed improved vector currents in $\Delta t/a_t = 16$ correlators\footnote{Source--sink separations, $\Delta t/a_t = 12, 16, 20, 24$ were computed to establish that suppressed excited--state contributions are under control. Details appear in the associated longer write-up.} averaged over ten source locations spread across the lattice.

\begin{figure*}
  \centerline{\includegraphics[width=.99\textwidth]{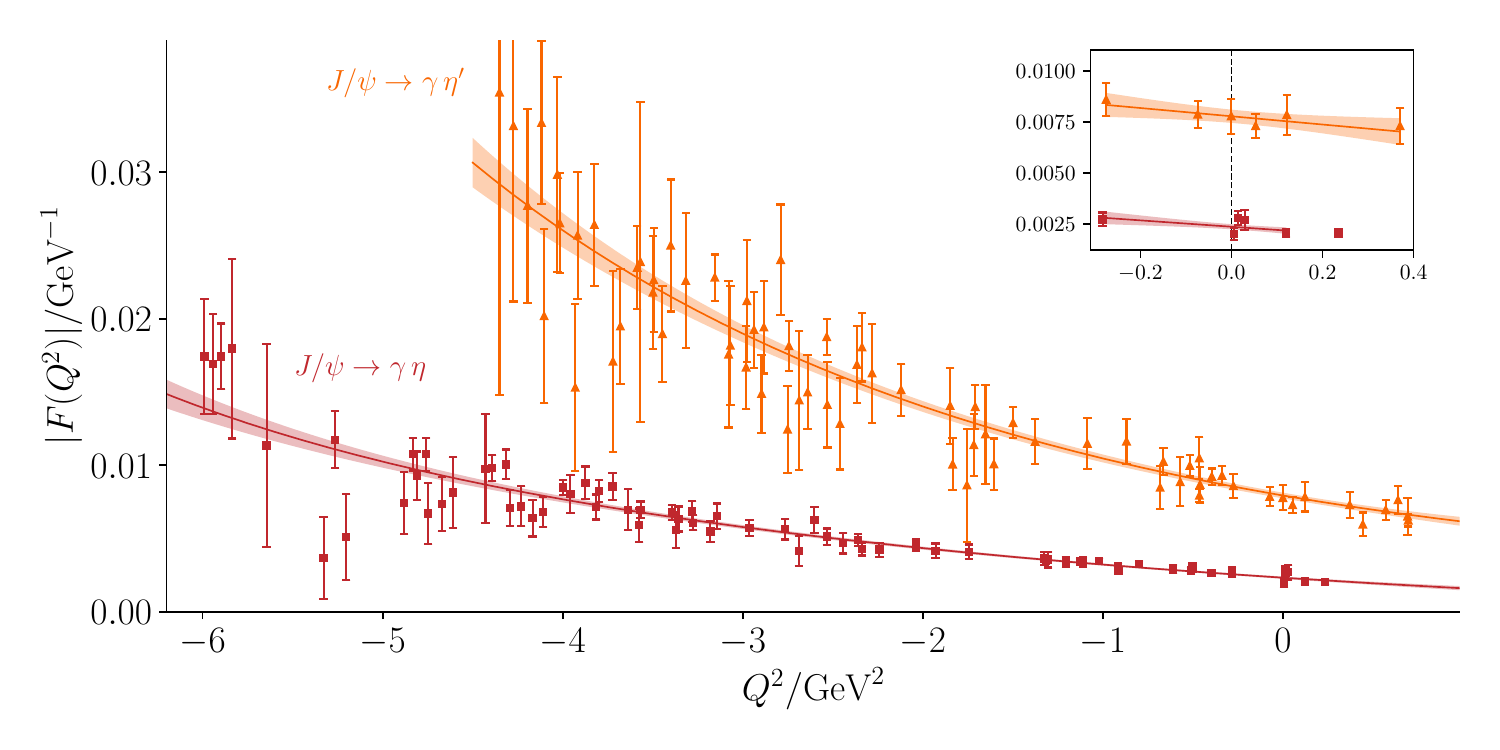}}
  \vspace*{-8mm}
  \caption[]{\label{fig:form-factors} Form--factors for $J/\psi \to \gamma \,\eta$ and $J/\psi \to \gamma \,\eta'$ as a function of photon virtuality, $Q^2$. Curves show one successful parameterization form as described in the text. The inset panel shows the region around $Q^2=0$, with bands showing a linear description of the data.
  }
\end{figure*}

\medskip
\pagebreak
\noindent\emph{Results} ---
Figure~\ref{fig:form-factors} shows the transition form--factors for $J/\psi \to \gamma \eta$ and $J/\psi \to \gamma \eta'$ determined using the technology described in the previous section. We observe clear consistency between extractions at neighboring $Q^2$ values, even though they may be obtained from correlation functions using rather different meson momenta. The relatively high statistical quality and consistency of data for the $\eta'$ show the power of optimized operators in extracting the excited--state contribution. The use of large meson momenta populates the region around $Q^2=0$ with several points, allowing a well--constrained determination of the real--photon transition rate via $|F(0)|$.
The systematically larger signal for $\eta'$ over $\eta$ is expected on phenomenological grounds: the $\eta'$ has a flavor structure that is dominantly $SU(3)_F$ \emph{singlet}, while the $\eta$ is dominantly \emph{octet}, with only a small admixture of singlet. In this radiative decay process, the annihilation of $c\bar{c}$ into a gluonic intermediate state should access only the \emph{singlet} components, leading to an enhanced production of $\eta'$, as we observe.

The data presented in Figure~\ref{fig:form-factors} had their $Q^2$ dependence described using fits to a number of parameterizations, with sufficiently flexible forms proving capable of doing so with $\chi^2/N_\mathrm{dof}$ near unity. 
The curves shown in Figure~\ref{fig:form-factors} correspond to quadratic--order polynomials in a conformal mapping of $Q^2$ to a variable in which convergence is expected to be rapid, away from the nearest expected singularity in these processes, which would be a pole at $Q^2 = - m_{J/\psi}^2$. These three--parameter descriptions can be seen to completely capture the $Q^2$--dependence of the data.

Description of the $Q^2$--dependence of $\eta$ production using a simple dipole form, 
$F(Q^2) = F_0 / (1+ Q^2/\Lambda^2)$,
as applied to experimental $J/\psi \to e^+ e^- \, \eta$ data in Ref.~\cite{BESIII:2018qzg}, and to the limited one-- and two--flavor lattice QCD data of Ref.~\cite{Jiang:2022gnd, Shi:2024fyv}, is of questionable quality, having ${\chi^2/N_\mathrm{dof} = 2.1}$. 
The value of $\Lambda$ found in the fit to the current data, \mbox{$2.596(14)$ GeV}, is quite compatible with that found in Ref.~\cite{BESIII:2018qzg}, $\Lambda_\mathrm{expt.} = 2.56(4)$ GeV.

Apart from the overall scale, the $Q^2$--dependences for the $\eta$ and $\eta'$ are compatible\footnote{For instance, description with $F_0 \, e^{-Q^2/16 \beta^2}$ gives ${\beta_\eta = 0.460(8)\, \mathrm{GeV}}$, ${\beta_{\eta'} = 0.463(10)\, \mathrm{GeV}}$ with $\chi^2/N_\mathrm{dof} \sim 1$ in each case.} supporting a plausible hypothesis that the production may straightforwardly factorize from the light meson dynamics in the final state, an observation which, if it continues to hold for more complex final states, may simplify analysis of light meson resonances produced in this process.

\begin{figure}
 \centerline{\includegraphics[width=.99\columnwidth]{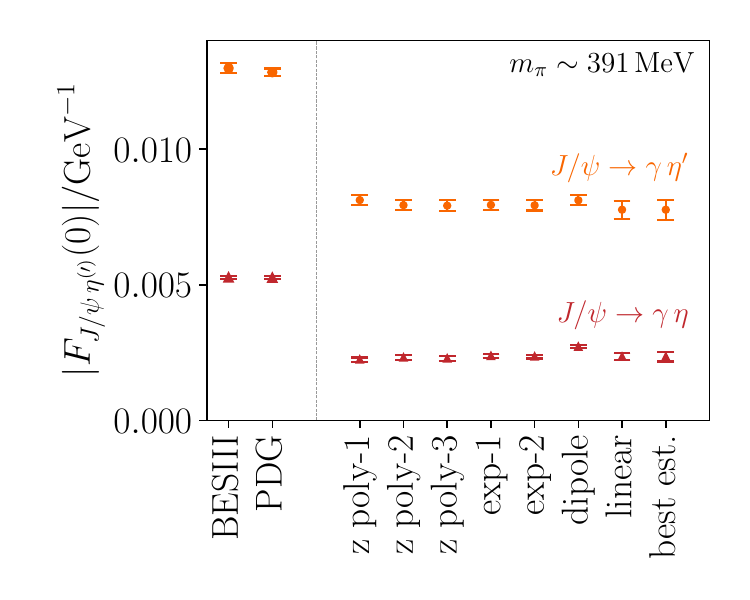}}
  \vspace*{-6mm}
  \caption[]{\label{fig:f0values} Form--factors for $J/\psi \to \gamma \,\eta$ and $J/\psi \to \gamma \,\eta'$ at ${Q^2 = 0}$, extracted from $Q^2$ parameterizations (discussed in detail in the associated longer write-up). The rightmost points are our best-estimates within the current calculation. Experimental values are provided on the left.
  }
\end{figure}

We determine the real--photon transition form--factors, $F(0)$, both from the parameterization fits to the $Q^2$--dependence, but also from linear interpolation of data points in the immediate vicinity of $Q^2=0$ (shown in the inset panel in Figure~\ref{fig:form-factors}). A summary is presented in Figure~\ref{fig:f0values} where we observe very little dependence upon the precise method of extraction\footnote{Except for the simple dipole fit to $J/\psi \to \gamma \eta$, which as discussed above describes the data somewhat poorly.}. Considering also the modest systematic error associated with variations in the anisotropy, $\xi$, we present final best--estimates at ${m_\pi \sim 391 \,\mathrm{MeV}}$ of
\begin{align*}
 \big| F_{J/\psi\, \eta}(0) \big| &= 0.00235(18) \, \mathrm{GeV}^{-1} \, ,\\
 \big| F_{J/\psi\, \eta'}(0) \big| &= 0.00777(37) \, \mathrm{GeV}^{-1} \, .
\end{align*}
In the same figure we also show the form--factor values as extracted from the current best measurement of the radiative decays by BESIII \cite{BESIII:2023fai} as well as the PDG average \cite{ParticleDataGroup:2022pth} (which includes the BESIII measurement).

From the best estimates in the current calculation, the ratio,
\begin{equation*}
\label{eq:2}
\frac{|F_{J/\psi\,\eta'}(0)|}{|F_{J/\psi\,\eta}(0)|} = 3.30(29) \, ,
\end{equation*}
which is somewhat larger than the value extracted from the PDG averages, $2.44(4)$.

\smallskip

The dependence of these form--factors on the mass of the light-quarks is not known, so any interpretation of these discrepancies with respect to experiment would be premature -- further calculations approaching the physical pion mass are required to establish the trend. Similarly, while we see no clear consequences of large discretization effects in the calculation, it is possible that the magnitude of these production amplitudes is sensitive to the lattice spacing.

\medskip
\noindent\emph{Outlook} ---
We have presented the first lattice QCD computation of the processes $J/\psi \to \gamma \, \eta$ \emph{and} $J/\psi \to \gamma \, \eta'$ determining both with good statistical precision over a range of photon virtualities for heavier than physical light quark mass.

The results of this calculation pose a mystery to be solved -- the magnitudes of both $\eta$ and $\eta'$ production lie well below the high precision experimental estimates. 
One explanation might be a strong light quark mass dependence, but there is no obvious reason why this should be particular to this process. Another option might be an imperfect realization of the topology of the QCD vacuum on this particular set of lattice configurations -- since the $\eta'$ mass, and presumably other properties of this state, are sensitive to the axial $U(1)$ anomaly (as is the $\eta$, but to a lesser extent) we might see sensitivity here. Indeed the somewhat unexpectedly small mass of the $\eta'$ on this lattice, $945(9)$ MeV, given the heavier--than--physical light quark masses, might be an indication that the anomaly contribution is under--estimated.

In summary, we have demonstrated a combination of technological advances that make large portions of the charmonium radiative decay sector accessible in lattice QCD computations. As well as further study of the $\eta$ and $\eta'$ production in $J/\psi$ radiative decays considered here, our progress suggests we can now pursue radiative decays of other charmonium states such as the $h_c$ and the $\psi(2S)$, as well as extension to the case of production of light meson \emph{resonances} in processes like $J/\psi \to \gamma\,  (\pi\pi, K\overline{K} \ldots)$, working towards the $J/\psi \to \gamma \, \eta \eta'$ process in which the exotic $\eta_1$ has been claimed.

%
\bigskip
\acknowledgments
We thank our colleagues within the Hadron Spectrum Collaboration for their continued assistance, and give particular thanks to C.E.~Thomas for his inspiration to consider the Wigner-Eckart averaging procedure. The authors acknowledge support from the U.S. Department of Energy contract DE-SC0018416 at William \& Mary, and contract DE-AC05-06OR23177, under which Jefferson Science Associates, LLC, manages and operates Jefferson Lab. 
This work contributes to the goals of the U.S. Department of Energy \emph{ExoHad} Topical Collaboration, Contract No. DE-SC0023598.
The authors acknowledge support from the U.S. Department of Energy, Office of Science, Office of Advanced Scientific Computing Research and Office of Nuclear Physics, Scientific Discovery through Advanced Computing (SciDAC) program. 
Also acknowledged is support from the Exascale Computing Project (17-SC-20-SC), a collaborative effort of the U.S. Department of Energy Office of Science and the National Nuclear Security Administration.

This work used clusters at Jefferson Laboratory under the USQCD Initiative and the LQCD ARRA project. This work also used the Cambridge Service for Data Driven Discovery (CSD3), part of which is operated by the University of Cambridge Research Computing Service (www.csd3.cam.ac.uk) on behalf of the STFC DiRAC HPC Facility (www.dirac.ac.uk). The DiRAC component of CSD3 was funded by BEIS capital funding via STFC capital grants ST/P002307/1 and ST/R002452/1 and STFC operations grant ST/R00689X/1. Other components were provided by Dell EMC and Intel using Tier-2 funding from the Engineering and Physical Sciences Research Council (capital grant EP/P020259/1). 

Also used was an award of computer time provided by the U.S.\ Department of Energy INCITE program and supported in part under an ALCC award, and resources at: the Oak Ridge Leadership Computing Facility, which is a DOE Office of Science User Facility supported under Contract DE-AC05-00OR22725; the National Energy Research Scientific Computing Center (NERSC), a U.S.\ Department of Energy Office of Science User Facility located at Lawrence Berkeley National Laboratory, operated under Contract No. DE-AC02-05CH11231; the Texas Advanced Computing Center (TACC) at The University of Texas at Austin; the Extreme Science and Engineering Discovery Environment (XSEDE), which is supported by National Science Foundation Grant No. ACI-1548562; and part of the Blue Waters sustained-petascale computing project, which is supported by the National Science Foundation (awards OCI-0725070 and ACI-1238993) and the state of Illinois. Blue Waters is a joint effort of the University of Illinois at Urbana-Champaign and its National Center for Supercomputing Applications.

The software codes
{\tt Chroma}~\cite{Edwards:2004sx}, {\tt QUDA}~\cite{Clark:2009wm,Babich:2010mu}, {\tt QUDA-MG}~\cite{Clark:SC2016}, {\tt QPhiX}~\cite{ISC13Phi},
{\tt MG\_PROTO}~\cite{MGProtoDownload}, and {\tt QOPQDP}~\cite{Osborn:2010mb,Babich:2010qb} were used.

\bibliographystyle{apsrev4-2}
\bibliography{biblio_long.bib}


\end{document}